\documentclass[%
twocolumn,
amsmath, amssymb
,aps
,prd
,floatfix
]{revtex4-2}

\usepackage{bm}
\usepackage{amssymb}	
\usepackage{amsthm}	
\usepackage{mathrsfs}
\usepackage[flushleft]{threeparttable}
\usepackage{amsmath}
\usepackage{tikz,scalerel}
\newcommand{\beq}{\begin{equation}}
\newcommand{\eeq}{\end{equation}}
\newcommand{\bea}{\begin{eqnarray}}
\newcommand{\eea}{\end{eqnarray}}
\newcommand{\ur}{\mathrm{U}}
\newcommand{\xb}{x_{_{Bj}}}

\begin{document}

\title{
Additional dimensions of space and time in the domain of deep inelastic processes
}%

 \author{B.~B.~Levchenko}%
\email{levchenko@www-hep.sinp.msu.ru }

\affiliation{%
D.V. Skobeltsyn Institute of Nuclear Physics, M.V. Lomonosov Moscow State University,\\ 
 Moscow 119991, Russian Federation\\
		}%

\begin{abstract}
We prove that the well-known Heisenberg uncertainty relations  and Landau-Peierls 
uncertainty relations implicitly contain   ``hidden'' angular variables,
which belong to  new uncertainty relations. 
Based on the obtained relations, we derive a formula  for estimating  
speed $U^*$ of a virtual particle in indirect measurements. 
We applied the theory of indirect measurements and the derived formula to estimate the module of the group velocity of virtual photons from the DIS HERA data.
The HERA data indicate that the speed of virtual photons  exceeds the speed of light $c$ in free space, $U^*>c$. 
The properties of virtual photons and a hypothetical tachyon particle are almost identical. 
It is found that in the realm of particle interaction, the new angular parameters are closely related to the type of the phase-space geometry and dimensionality of the spacetime continuum.
It is suggested that  the problem of the normalisation condition $U^* =c$  at $Q^2=0\, \rm{GeV}^2$
can be solved naturally   within the  framework of  ``Two-Time Physics'' developed by I. Bars.  2T-physics is the theory   with  local symplectic $\mathrm{Sp(2,R)}$ gauge symmetry in phase-space and the spacetime geometry  of  signature $\mathrm{(1+1',d+1')}$ with  one extra time-like and one extra space-like dimensions. 
\end{abstract}

\maketitle

\section{Introduction}
Physical processes in quantum systems can only be properly described by  introducing various  types  of intermediate states.  
Scattering processes of real particles are also described by the quantum mechanism of
exchange  of virtual particles (gauge bosons, resonance states, more complex objects, such as Regge trajectories). 
The concept of  a virtual off-mass-shell
particle is derived from the microscopic violation of causality allowed
by the time-energy uncertainty relation \cite{WH, LP31, Heitler, BLP, HKN}.
The relation between the momentum and energy of a virtual particle can be anything that is required by the conservation of 4-momentum at the vertices.
It should be noted that the content of the term ``virtual particle'' has undergone a significant change. Even in the recent past, virtual particles usually meant such particles in virtual states (e.g.  photons, electrons, pions) that were well studied in real states. A class of particles (quarks, gluons, etc.) has emerged which, due to the confinement property of quantum chromodynamics, cannot in principle be in real states. 
 
Although it is impossible to  observe such intermediate states directly, their experimental study is of great interest and importance due to their nontrivial dynamical properties.
Nevertheless, a number of  properties of the virtual particle can be measured indirectly.
Indirect measurement is a measurement in which the value of the unknown quantity sought is calculated from  measurements of other quantities related to the measurand by some known relation \cite{M72, Rab05}. 

The  goal of the  present communication is twofold.
First, we derive a formula for estimating a speed of virtual particles. 
Second,  we combine the theory of indirect measurements, the hardware resolution of the ZEUS detector
 \cite{zeus_ctd, zeus_cal}  and   the obtained formula 
into a mathematical tool for evaluating  the speed of virtual particles
from  experimental data. As an application,
we present  preliminary results on the speed of virtual photons ($\gamma^*$)
using a small set of deep inelastic scattering (DIS)
data from the HERA collider \cite{DESY_15_039}. 

Natural questions arises: What is the value of data on the speed of virtual particles?  
What new information would they reveal?
Especially considering that
at relativistic velocities, a particle's speed  is less informative than its momentum. And in the case of a real photon, knowing only its speed tells us nothing about the photon's energy or momentum.

Let us recall three phenomena: 1) The formation of a Mach cone when the speed of a body in a medium  is close to  or greater than the speed of sound in the medium; 
2) The Cherenkov-Vavilov effect \cite{Ch34, Ch37,W34};
 3) In 1904-1905 A. Sommerfeld \cite{Som1905_2} 
established   in the context  of Lorentz's theory of electromagnetism \cite{L1904} that if the speed of the 
electron $\ur$ is less than the speed of light in a vacuum, $\ur <c$, then the electron is able to move   at a constant velocity. However, if  $\ur> c$, then  an external force is required for uniform motion. 
It is not superfluous to note that most of these conclusions were made much earlier by O. Heaviside 
\cite{H1889, Tyap74}.

What these examples have in common is that when  the particle speed exceeds a certain characteristic value, $ \ur_{cr} $, the  dynamics  of the process  changes dramatically. Moreover, the presence of particles with velocities above the critical value serves as a natural indicator of changes in the properties of the phase space itself in which the process is taking place.

The ``implementation'' of the uncertainty principle in nature determines the existence of virtual particles with  a very wide  range of dynamic properties. 
For this reason, the present  derivation of a formula for the  lower bound on the
speed of a virtual particle is based on the Heisenberg's uncertainty relations (HUR).

The uncertainty principle and the uncertainty relations for observables of two canonically conjugate quantum mechanical operators discovered by Heisenberg  
\cite{WH} are fundamental foundations of quantum mechanics.  
In the article \cite{WH}, only a heuristic estimate was given of how the inaccuracy of  the particle coordinate, $ q_1 $,  is associated with the inaccuracy of the particle 
momentum, $ p_1 $, into one relation, $ p_1 q_1 \sim \hbar$, called the uncertainty relation
( see also Refs \cite{Kennard27}, \cite{Rob},  \cite{Schrd}). 
H. Weyl  \cite{Weyl_e} provide   another proof of HUR   and also gave the inequality a modern look,
\beq
\Delta p_x  \Delta x\geq \hbar/2.
\label{HUR}
\eeq

In the next two sections
we present derivations of new uncertainty relations and a formula for evaluating the speed of a virtual particle in indirect measurements. 
After presenting experimental results for the lower bound on the speed of virtual photons, we discuss the tachyonic properties of $\gamma^*$, followed by arguments that allow relating the superluminal speed of virtual photons to the metric properties of spacetime (in particular, its dimensionality).

\section{New  angular variables and uncertainty relations}
Uncertainties of quantum mechanical
Hermitian operators $\hat{x}$ and $\hat{p_x}$ are defined 
(see Ref. \cite{Weyl_e}, p. 77 and  Ref. \cite{Griff}, p. 137) via
\beq
(\Delta x)^2={\mathrm \int^{+\infty}_{-\infty}}\!\! x^2{\bar{\varphi}}\varphi{ d}x,\,\,\,
(\Delta p_x)^2=\int^{+\infty}_{-\infty}\!\! {\bar{\varphi}}\frac{{\partial}^2\varphi}{{\partial}x^2}{ d}x.
\eeq
Therefore, we write out the uncertainty relations for all projections of the pair of conjugate coordinate-momentum pair  in terms of mean square deviations:
\bea
(\Delta p_x)^2 (\Delta x)^2&\geq& (\hbar/2)^2, \nonumber \\
(\Delta p_y)^2 (\Delta y)^2&\geq& (\hbar/2)^2, 
\label{3ur}\\
(\Delta p_z)^2 (\Delta z)^2&\geq& (\hbar/2)^2, \nonumber
\eea
without extracting the square root, as in (\ref{HUR}).
If we now add the left-hand sides of these inequalities, 
we find that the resulting  sum is a dot product 
$\mathbf{(\Delta P)^{(2)} \cdot (\Delta R)^{(2)}}$
of  the vectors 
$\mathbf{(\Delta P)^{(2)}}=((\Delta p_x)^2,(\Delta p_y)^2,(\Delta p_z)^2)$ 
and
$\mathbf{(\Delta R)^{(2)}}=((\Delta x)^2,(\Delta y)^2,(\Delta z)^2)$. 
The dot product specifies the angle between vectors, and therefore for the 
norm of  vectors $\mathbf{(\Delta P)^{(2)}}$ and $\mathbf{(\Delta R)^{(2)}}$ 
we obtain the uncertainty relation, which includes a new angular variable $\psi$:
\beq
\Vert \mathbf{(\Delta P)^{(2)}}\Vert\, \Vert \mathbf{(\Delta R)^{(2)}} \Vert\geq \frac{3\hbar^2}{4\cos \psi}.
\label{svp}
\eeq

The relations (\ref{3ur})  consist of  only positive definite terms and this defines  the domain of the angle  $\psi\,\in[0, \pi/2)$, and  the domain of the function values, $0<\cos {\psi} \le 1 $. 
Thus, depending on the state of the  physical system under study,  the value of $\cos{\psi}$
 varies and imposes constraints  on $\Vert \mathbf{(\Delta P)^{(2)}}\Vert$ and 
$\Vert \mathbf{(\Delta R)^{(2)}}\Vert$ of different degrees of stiffness.
The function $\cos{\psi}$  appears as a result of reducing the six degrees of freedom in  (\ref{3ur}) to three  degrees of freedom in (\ref{svp}).  

\section{Estimation of a particle speed from the uncertainty relations}
In the same  1927 article \cite{WH}, Heisenberg gives an uncertainty relation for 
another pair of canonically conjugate energy-time variables. This relation is only definite  up to Planck's constant, so we write it out by including an arbitrary constant 
$\delta_H $:
\beq
(\Delta E)^2 (\Delta t)^2 \geq {\delta^2 \!\!\!}_{_H}\,\hbar^2 ,
\label{HB1}
\eeq
whose value is fixed by the conditions of the problem being solved. 
Landau and Peierls  \cite{LP31}  generalized a number of conclusions from classical 
quantum mechanics to the relativistic domain. In particular, it was shown  that 
the Heisenberg inequalities for momentum and coordinates are also valid at relativistic  velocities. 
In passing to the relativistic consideration, however, the inequality (\ref{HB1}) 
does not  give such a simple justification. 
Nevertheless, Landau and Peierls have 
derived  new inequalities for a free relativistic particle,
 Refs. \cite{LP31, FK47}:
\beq
|\ur_i|\Delta p_i \Delta t \ge {\delta\!}_{_{LP}} \hbar,
\label{LP}
\eeq 
that holds for each of the components $i=(x,y,z)$ 
separately. 
Here the symbol $\mathbf {U}$ denotes the group velocity vector of the particle, 
$\mathbf {U}=(\ur_x, \ur_y, \ur_z)$ and 
an arbitrary constant $\delta_{_{LP}} $ is introduced on the same 
reasoning as in the inequality (\ref{HB1}).
Adding the squares of the relations (\ref{LP}) for $ i = (x, y, z) $, 
as above, we get on the left-hand side of the inequality the scalar product 
$\mathbf{U^{(2)} \cdot (\Delta P)^{(2)}}$
of   vectors $\mathbf{U^{(2)}}=((\ur_x)^2,(\ur_y)^2,(\ur_z)^2)$ and
$\mathbf{(\Delta P)^{(2)}}$. 
 Thus we  reveal another  ``hidden'' angle  ${\psi}_{_{H}}$ between the phase-space vectors  and obtain another  inequality connecting the norms 
of  the particle's  quadratic velocity vector, the mean square deviation of 
its momentum and the square of the duration of the measurement process:
\beq
\Vert\mathbf{U^{(2)}}\Vert\,\Vert \mathbf{(\Delta P)^{(2)}}\Vert(\Delta t)^2
\geq 3(\delta_{_{LP}}\hbar)^2 /\cos {\psi}_{_{H}}.
\label{LPL}
\eeq

Using inequalities (\ref{svp}) and (\ref{LPL}), we are now able to  estimate the module of the particle group velocity  $|\mathbf{U}|$ in conditions where the direct measurement of the velocity is impossible (the method of indirect measurements  \cite{M72}).
For this purpose,  the ratio  of the inequality (\ref{LPL}) to (\ref{svp}) 
or the  ratio  of the inequality (\ref{LPL}) to (\ref{HB1}), respectively, must be taken.
In this way,
\beq
\Vert \mathbf {U^{(2)}}\Vert\geq A_t\frac{(\Delta E)^2}{\Vert \mathbf{(\Delta P)^{(2)}}\Vert}.
\label{Vel2}
\eeq

Finally, by means of the Cauchy-Buniakowsky-Schwarz inequality, we obtain the following estimate of 
{\it the lower bound} of the norm of the velocity,
\beq
\Vert \ur_{lb}^* \Vert\sim \sqrt{\sqrt{3}\Vert \mathbf{U^{(2)}}\Vert}=\sqrt{\sqrt{3}A_t\frac{(\Delta E)^2}{\Vert \mathbf{(\Delta P)^{(2)}}\Vert}}.
\label{Vel_lb}
\eeq
Here $A_t= 3{\delta^2\!\!\!}_{_{LP}}/({\delta^2\!\!\!}_{_H}\, \cos {\psi}_{_{H}})$ is the  theoretical magnitude of the normalisation parameter. 
In the next section, we will see how the velocity of virtual particles relates the value of $A_t$  and the metric properties of spacetime.

\section{Spacetime metrics in the interaction domain}
To classify (pseudo-)Euclidean spaces, the so-called space index $ k $ (or the index of inertia) is 
introduced. It is defined as the number of imaginary unit basis vectors of the 
orthonormal frame \cite{Rash67, Shafarevich13}. For the proper Euclidean space, $n=d$, the space index  $ k = 0 $. For   the Minkowski space with the  total dimension $n =d+1= 4$ and the signature (+,-,-,-)=(1, 3), the space index of $k=3$.

\begin{figure*}
\centering
\begin{minipage}[t]{.495\textwidth}
\includegraphics[width=\textwidth]{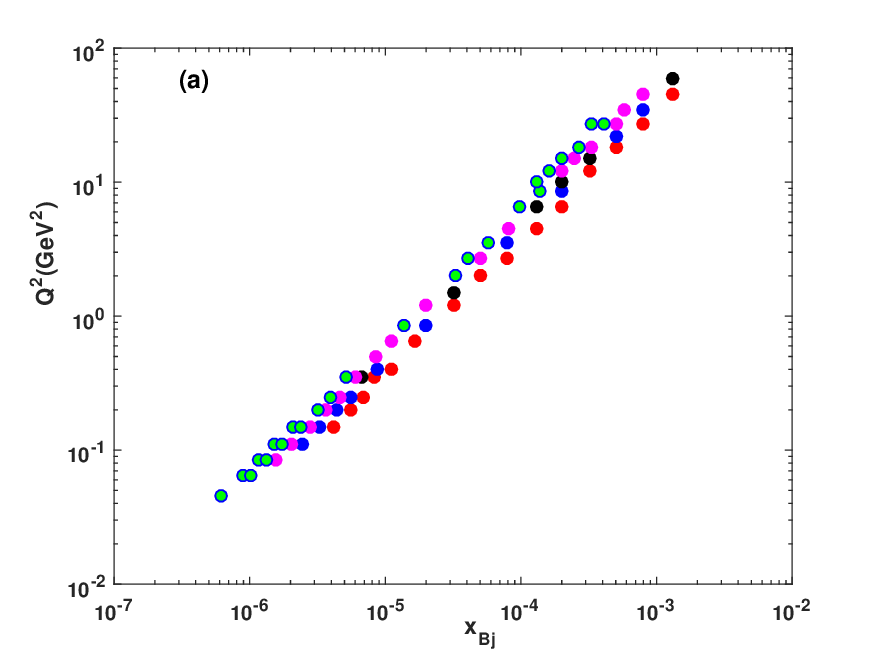}
\end{minipage}	
\begin{minipage}[t]{.495\textwidth}
\includegraphics[width=\textwidth]{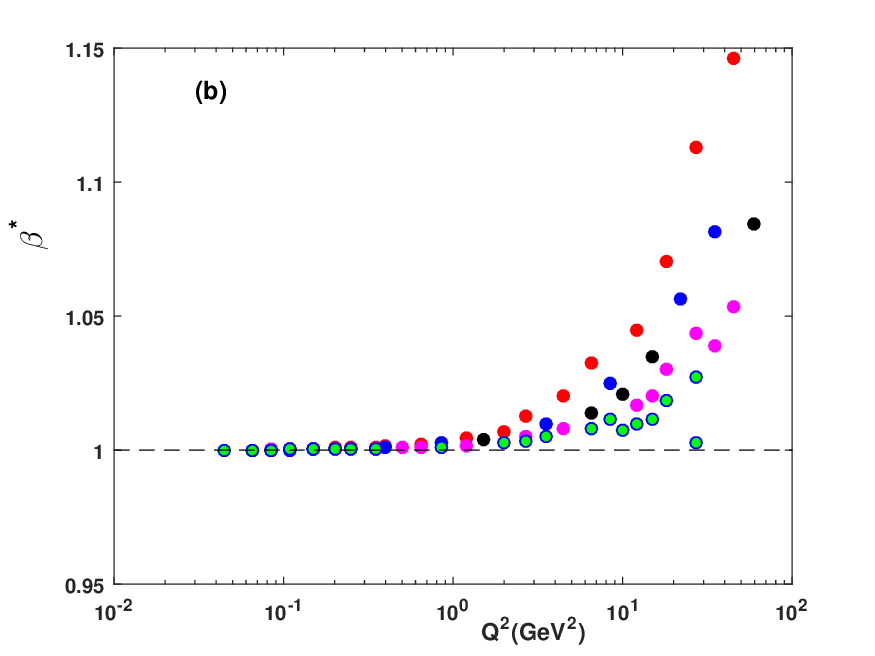}
\end{minipage} 
\begin{minipage}[t]{.495\textwidth}
\includegraphics[width=\textwidth]{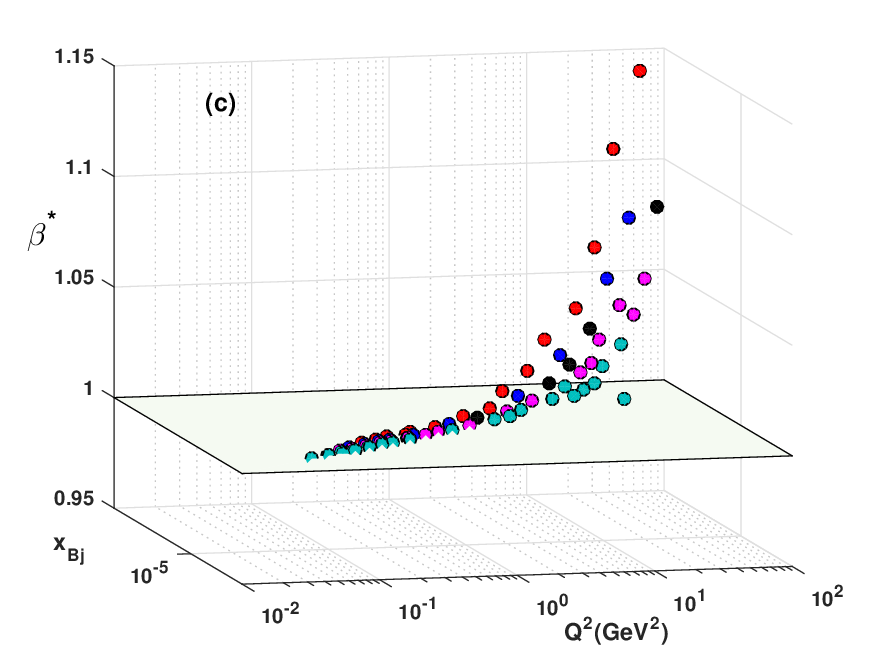}
\end{minipage}
\begin{minipage}[t]{.495\textwidth}
\includegraphics[width=\textwidth]{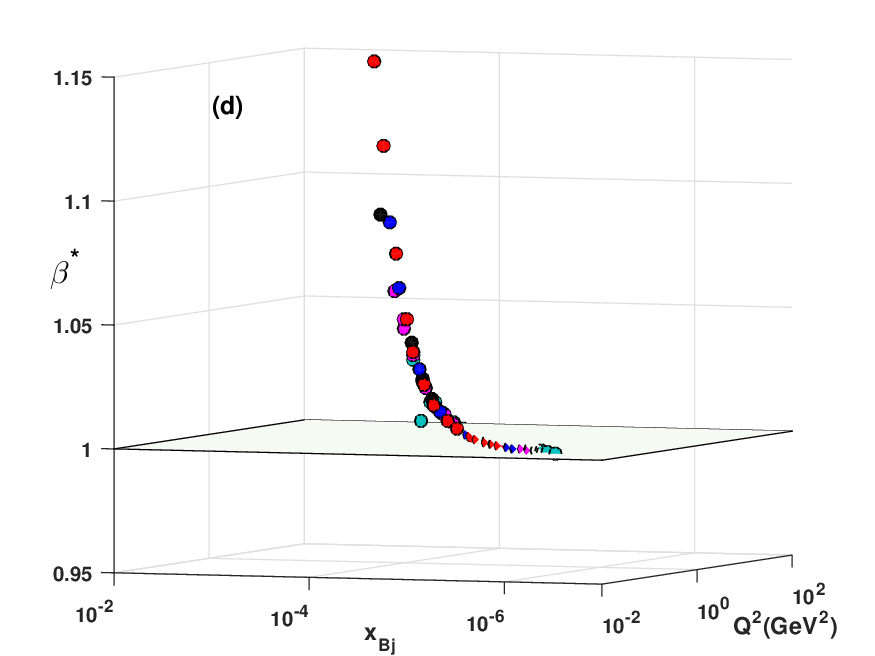}
\end{minipage}
\noindent
\caption{ (a) Kinematic  ($\xb$, $Q^2$) grid of
the combined HERA I   data  for neutral current  $e^+p$ 
 deep inelastic scattering events at the center-of-mass energy 
 $\sqrt{s_{com}}\approx 300$ GeV   \cite{DESY_15_039}, Table 11. 
Different colors  indicate data with the inelasticity, $y=Q^2/\xb{s}_{com}$,
 in the following intervals 
: $0.354-0.47$ (red);  $0.47-0.511$ (blue); $0.511-0.598$ (black); $0.598-0.676$ (magenta);
$0.676-0.951$ (green).
(b) The normalized virtual photon speed, $\beta^* = \Vert \ur_{lb}^* \Vert/c$, 
as a function of the photon virtuality $Q^2$.  
The color of the dots corresponds to different values of the inelasticity $y$ as in Fig (a).
It follows from the condition, 
$\beta^*(Q^2\rightarrow 0)=$1 that the  normalisation  parameter $A_e=1/\sqrt{3}$.
(c) The same as in panel (b), but for $\beta^* $ as a function of  two  variables $\xb$ and $Q^2$ at
different $y$-intervals.
(d) The same as in panel (c), but  the figure is rotated so that the data points are roughly projected onto a curve.}
\label{fig1} 
\end{figure*}

Let us first discuss  possible values of the parameter $A_t$ in formula (\ref{Vel_lb}), 
and denote by $A_e$ its value found from the data.
The Standard Model assumes that the geometry of spacetime known from macroscopic physics also holds in the microcosm too.
The derivation of the inequality (\ref{svp}) was based on this assumption,  using  the dot product of vectors in proper Euclidean space.
The inequality (\ref{LP}) was derived from the  inequality (\ref{HB1}). Therefore,
there is  a good reason to believe that 
${\delta^2\!\!\!}_{_{LP}}={\delta^2\!\!\!}_{_H}$. 
In this case,  $ A_t = 3 / \cos {\psi} _ {_{H}}\ge 3 $. 
Consequently, if it follows from an experimental data that $ A_e \ge 3$,
then the interaction of particles takes place in the domain of spacetime with the Minkowski  geometry and the space index  of  $k=3$. 

The case is quite different if  $A_e <3$. Then our assumption about the metric of spacetime in the interaction domain is not correct.

Let us now apply the formula (\ref{Vel_lb}) to evaluate  the speed of  virtual photons.
In DIS processes  $c|\vec{q}|=\sqrt{q_0^2 + Q^2}$
and therefore the quantities   $(\Delta E)^2 =(\Delta q_0)^2$ 
and $\Vert \mathbf{(\Delta P)^{(2)}}\Vert = (\Delta q)^2$  
depend on the kinematic variables  $\xb$, $y$, $Q^2$ and   uncertainties of their measurements
by the following chain of relations,
\begin{eqnarray}
(\Delta q)^2 &=& \frac{1}{c^4}\Big [ \Big ( \frac{ q_0 }{ |\vec{q}|} \Big )^2 (\Delta q_0)^2 
+ \frac{(\Delta Q^2)^2}{4(\vec{q})^2}  \Big ],\label{un_p} \\
(\Delta q_0)^2 &=& c^2P^2y^2 (\Delta \xb)^2\nonumber \\ 
&& +\,c^2 (l-\xb P )^2(\Delta y)^2, \label{un_e} \\
(\Delta Q^2)^2 &=& \frac{4 Q^2}{1-y} (c\Delta p_t)^2
+  \Big (\frac{ Q^2}{1- y } \Big )^2(\Delta y)^2,\label{un_q2} \\
(\Delta \xb)^2 &=& \frac{4 \xb^2}{(1-y) Q^2 } (c\Delta p_t)^2\nonumber \\
&& +\,\xb^2\Bigg [\frac{1}{(1- y)^2 }+\frac{1}{y^2 } \Bigg ](\Delta y)^2.\label{un_xb}
\end{eqnarray}
The chain of these relations is closed if to  enter the resolution of the central tracking  detector
 $\sigma(p_t)/p_t$ and the energy resolution of the uranium calorimeter $\sigma(\Sigma_e)/\Sigma_e$ 
\cite{zeus_ctd, zeus_cal}.

As input, we use the combined data from the H1 and ZEUS experiments
 on deep inelastic $ ep $ scattering at the HERA collider \cite{DESY_15_039}.
The magnitude of the  exchange particles virtuality, $ Q^2 $,  varies over a very wide range  of values. 
Neutral current  interaction cross sections at low $Q^2\le $100 GeV$^2$
are dominated by the virtual photon exchange. 
In the limit $ Q^2 \rightarrow 0$ GeV$^2$ (a real photon limit), 
$\beta^*=\Vert \ur_{lb}^*/c \Vert \rightarrow 1$
should hold.  This  condition allows to fix  the value of  $ A_e $
and at $ Q^2 >0$ GeV$^2$ to set {\it a lower bound} on the speed of virtual photons.

The kinematic range of  the combined HERA I data with $Q^2\le $100 GeV$^2$   is shown  in Fig. 1(a). 
Data points are grouped into strips with similar  values of inelasticity $y$ and marked with different colors (see figure caption). Such structuring reflects the kinematic relationship between the variables $\xb$, $y$ and $Q^2$, and the procedure for combining data from two different experiments by translation onto common grids \cite{DESY_15_039}.

Figure~1(b) shows the  speed of virtual photons normalized to the speed of real photons, $\beta^* = \Vert \ur_{lb}^* \Vert/c$,   as a function of  $Q^2$  at different $y$-intervals. 
The result follows from Eqs. (\ref{Vel_lb}) - (\ref{un_xb}) and
the HERA I data   for neutral current  $e^+p$  deep inelastic scattering events 
with beam  momenta $(l, P)=$(27.5, 820) GeV/c and
the center-of-mass energy $\sqrt{s_{com}}\approx 300$ GeV    \cite{DESY_15_039}, Table 11. 
Note that $\beta^*$ grows with $Q^2$, but the slope of this growth decreases with $y$.
The condition $\beta^*(Q^2\rightarrow 0)=$1 fixes $A_e$. In this way one get $\sqrt{3}A_e=1$.
Figures 1(c) and 1(d) show the dependence of $\beta^*$  on two variables, $\xb$ and $Q^2$. And again we see that  $\beta^*$ grows with $Q^2$ and $\xb$.
Figure 1(d)  is rotated view of Fig. 1(c) in order to project all data points on the same curve. This demonstrates that almost all points at different $y$ are located on a flat surface. 
  
The results presented in Fig.~1 show that the speed of virtual photons at 
$ Q^2 > 0$ GeV$^2$ exceeds the speed of light $c$ in free space, $\beta^*>1$.
A perusal of the litrature on faster-than-light  particles reveals that 
 virtual photons can be interpreted as  representatives of a class of superluminal particles, hypothetical tachyons   \cite{tach_BS62,  tach_F67, tach_BS69,  Terl68, Barashenkov1975, tach_S}, since their properties are largely similar as shown in Table~\ref{tab:table1},
\begin{table}[!h]
\caption{\label{tab:table1} Comparison of the properties of  virtual photons  and tachyons.}
\renewcommand{\arraystretch}{1.3}
\begin{center}
\small  
\begin{tabular}{@{}l c c }
\hline\hline
 &Virtual photon, $\gamma^*$   & Tachyon\\
\hline
mass    & $(m^*)^2=-Q^2<0$ &$(m^*)^2<0$ \\
energy &$q_0 <0, >0$         &$\epsilon <0, >0$\\
speed  & $\beta^* \ge 1$       &$\beta^* \ge 1$ \\
 &$q_0 \rightarrow 0, \beta^*  \rightarrow \infty$ (?) &$\epsilon \rightarrow 0, \beta^*  \rightarrow \infty$\\
\hline\hline
\end{tabular}
\end{center}
\end{table}
Moreover, such a result has been expected for a long time. Thanks to many years of extensive theoretical studies, mainly carried out by a group of Italian physicists led by E. Recami 
(for an overview see \cite{R_rev1, R_rev2}), it became clear that tachyons (or spacelike states) must exist as intermediate states or exchanged objects in elementary processes 
\cite{R_is1, R_is2, R_rev1}.
For reviews of searches for evidence of tachyons see \cite{R_rev3, tach_E}.

Thus, using a different theoretical approach and based on modern experimental data from high-energy physics, we have confirmed the predictions made more than 40 years ago.

From the above discussion, the value of $A_e=1/\sqrt{3}< 3$  obtained from the experimental data,  contradicts the theoretical expectation,  $ A_t = 3 / \cos {\psi}_{_{H}} > 3 $.  
Therefore, we want to identify a theoretical scenario that can help  eliminate this contradiction.

 I) {\it Nonlinear fields in 4D} \cite{ DBloOrl53, DBlo73}. 
In  nonlinear field theories (with scalar or/and vector fields) the signal velocity 
depends on the magnitude of the field and its derivatives. In this case, the speed 
of the signals can be less or more  than the speed of light $ c $ in void and the space index change from $k=3$ to $k=2$ or even to $k=1$ and hence, $A_t< 3$. 
This corresponds to the spacetime domain, in which there are two or three time-like dimensions. However, in this scenario there are problems not only with causality (paradoxes),  the violation of unitarity and occurrence of  ghost modes with negative norm (for more discussion see Refs. 
\cite{Bars_Tern, J_Muell}), but also the entire relativistic kinematics in the Minkowski space is destroyed.

II) {\it Stochastic time.} One can  increase the pseudo-Euclideanity 
of spacetime by adding an additional time-like dimension in the interaction domain, i.e., going from signature (1,3) to signature (2,3).
Analysis of this scenario shows \cite{J_Muell}, that in the case of  a spacetime
 with a thermally excited second time dimension the dynamics of the physical system is diffusive, not ballistic, so that all trajectories involving motion in the second
time dimension are dynamically unstable,  thus allowing us to avoid the 
difﬁculties outlined for nonlinear field theories.
The thermal extra time dimension behaves like an extra spatial dimension and 
the high temperature limit for the thermal extra time dimension is equivalent to a
small compactiﬁcation radius for an extra spatial dimension. 
  Therefore, such a scenario is consistent with Kaluza-Klein type theories.
However,  simply adding another  spatial or temporal dimension  does not solve the problem we  encountered ($A_t>3$, $A_e<3$).

III) {\it Two-time physics.} Now we have to account that the vectors $\mathbf{X}$, $\mathbf{P}$ and 
$\mathbf{(\Delta R)^{(2)}}$,  $\mathbf{(\Delta P)^{(2)}}$  are elements of the phase space. 
In the geometric approach, the phase space  $(\mathbf{Q}^i, \mathbf{P}_i)$ of classical mechanics 
as well as of quantum mechanics is taken to be a smooth manifold equipped with a symplectic form, which induce  a so-called symplectic geometry \cite{KimN91}. 
Note that in the early 60's attempts were made to systematize hadrons with the use of the symplectic group Sp(6) \cite{VanHove65, Strum65}, and in the 80's gauge theories with Higgs models based on simple classical Lie groups \cite{Vol89} and, in particular, the  symplectic group Sp(m) were investigated. 

The principle of local gauge invariance is an important
ingredient in the construction of realistic models for the
interactions observed in nature. I.~Bars  with co-authors \cite{BarsAnDD},
\cite{Bars01}, \cite{Bars_Tern}
discovered a fundamental role of the local symplectic Sp(2, R) gauge symmetry
in phase space, which gave rise to new field theories  in (2,4) spacetime with one extra time and one extra   
space  dimensions (2T-physics). This includes 2T field theories that
yield one-time (1T)  field theories for the Standard Model  \cite{BarsSM}, General Relativity \cite{BarsGR},
SUSY \cite{Bars2010zw} and others. 
The canonical transformations of the type Sp(2, R),  considered as a local gauge symmetry,
has the power to cure the ghost and causality problems of extra time-like dimensions.

In the following, we give a qualitative argument for the fact that the condition $A_t<3$ is feasible in the framework of 2T-physics.
First, we postulate that  the interaction domain of DIS has a small spatial and temporal extent, and
within the framework of the 2T-physics, the interactions  take place in  $\mathrm{(1+1', 3+1')}$ space 
with one extra time  and one extra space dimensions and 
the gauge invariant sector of 2T-physics, namely the ghost free physical sector, effectively becomes  a
  1T theory with {\it an effective} 1+3 dimensions 
	\cite{BarsGS10}.
Mathematically, the  speed of a photon in the physical sector
 is a combination of the three effective spatial dimensions and one effective 
temporal dimension (the "effective Minkowski spacetime").
So, we assume that during the DIS process a virtual photon ``captures'' (or ``perceives")  the ``new'' additional spatial $x$ or  time-like $\tau$ dimensions with the probability $\alpha$
and  captures  the ``old'' spacetime dimensions, $\{ t, x^i \}$, with the probability  
$\omega$.  We label the components containing the conventional time dimension with 
index $H$, and the components containing an additional time-like dimension with index $B$.
As a result, the squares of the normalized photon velocities as spatio-temporal combinations give contributions with the following probabilities:
${\beta^2}_H(t,x^1,x^2,x^3)\sim \omega$,
$\tilde{\beta^2}_H(t,x^i,x^j,x)\sim 3\alpha$, ${u^2}_B(\tau,x^1,x^2,x^3)\sim \alpha$ and 
$\tilde{u^2}_B(\tau,x^i,x^j,x)\sim 3\alpha^2$.
The total probability must satisfy the condition:
\bea
\omega + 4\alpha + 3\alpha^2 =1.
\label{alfa}
\eea
In particular, $\alpha = 0$ if $\omega = 1$, in accordance with the definition of the introduced probabilities.

Let us now find out how the parameter $A$ changes when additional dimensions are taken into account, and denote it  as $A_t = A_{eff}$. 
Then, in accordance with Eq. (\ref{Vel_lb})  one get
 \bea
{\beta^*}^2_{eff} =\sqrt{3}(\omega{\beta^2_H}A_H+3\alpha\tilde{\beta}^2_H A_H \nonumber\\
+ \alpha{u^2_B} A_B + 3\alpha^2 \tilde{u}^2_B A_B).
\label{beff}
\eea
Here, $A_H= 3/ \cos {\psi}_{_H}$ and $A_B= 3/ \cos {\psi}_{_B}$.

In the limit $Q^2\to 0$, all speeds  in Eq. (\ref{beff})
should be equated to unity. As a result, we get the normalisation condition 
$\sqrt{3}A_{eff}=1$, 
 \bea
1=3\sqrt{3}\Big [\frac{1-\alpha - 3 \alpha^2}{\cos {\psi}_{_H}}    
+\frac{\alpha  + 3\alpha^2}{\cos {\psi}_{_B}}\Big].
\label{Aeff}
\eea
By solving this equation for $\alpha$, we find that $\alpha > 0$ if 
$\cos {\psi}_{_B}> \cos {\psi}_{_H}$ and $\alpha < 1$ 
if the following condition is met
\beq
(12\sqrt{3} - \cos {\psi}_{_B})\cos {\psi}_{_H}  < 9 \sqrt{3}\cos {\psi}_{_B} \nonumber.
\eeq
These restrictions on $\cos {\psi}_{_H}$ and $\cos {\psi}_{_B}$ are quite soft.
Thus, in the spacetime volume of a high-energy reaction a ``mixing in'' of extra dimensions is possible and  the normalisation condition $\sqrt{3}A_{eff}=1$ is feasible.
The specific values of $\cos {\psi}_{_H}$, $\cos {\psi}_{_B}$, $\alpha$, ${\beta^2}_H$, 
$\tilde{\beta^2}_H$, $\tilde{u^2}_B$, $\tilde{u^2}_B$ 
 can be found by fitting Eq. (\ref{beff}) to the data similar to  Fig. 1(c).

\section{Conclusions}
This communication presents formula (\ref{Vel_lb}) for estimating the lower bound of the modulus of the group velocity of a virtual particle.
When analysing experimental data, the uncertainties included in the formula should be calculated by methods of indirect measurement theory. 
To estimate the speed of  virtual photons, the combined data from the H1 and ZEUS experiments on deep inelastic $ep$ scattering at the HERA collider were used as input.
The HERA data at  $Q^2\le $100 GeV$^2$ show that  the normalized speed of virtual   photons, $\beta^*$, exceeds  the speed of light in free space, $\beta^*\ge 1$.
The superluminal speed means that the virtual photons $\gamma^*$ behave  tachyon-like.
This characteristic behavior of virtual photons is consistent with the predictions made by Recami and co-authors in the 1970s and 1980s that tachyons are  intermediate states or exchanged objects in elementary processes.
 
When the normalisation condition $\beta^* =1$  is imposed  in a real photon limit, 
the deep relationship between the normalisation parameter $A_t$ and the type of  spacetime 
geometry in the interaction domain is revealed.
A solution of the problem of the normalisation condition $\beta^* =1$  at $Q^2=0\, \rm{GeV}^2$ is proposed in the framework of  ideas  of "Two-time physics" developed by I. Bars.
It is shown that by admixing one extra time and one exta space dimensions to 4D spacetime in the domain of DIS processes, it is possible to satisfy the normalisation condition $\sqrt{3}A_{eff}=1$ as observed in the effective physical (1,3) Minkowski spacetime.

For the sake of illustration of the method, the results for virtual photons presented here are based on only a small fraction of the HERA data, but the analysis can be extended to the full data set, and these investigations are left to future work.

\section*{ACKNOWLEDGMENTS}
The author  thanks M. Wing for reading an earlier version of  the manuscript and 
suggestions, E. Boos, L. Dudko, A. Geiser, N. Nikitin, E. Oborneva,  I. Volobuev  
and other colleagues from SINP MSU and the ZEUS collaboration for discussions and comments.
The author is especially grateful to  K. G. Gulamov, N. N. Nikolaev and B.S. Yuldashev for 
their mentorship in the early days of this  amazing scientific journey. 
The author expresses his gratitude to the Ministry of Science and Education of the Russian Federation and the DESY Directorate for their long-term financial support and for the hospitality extended to the non-DESY members of the ZEUS collaboration.

\bibliography{hiddennn}
\end{document}